\documentclass[preprint,12pt]{elsarticle}

\usepackage{graphicx}

\usepackage{amssymb}

\begin{document}

\begin{frontmatter}



\title{Is SN 2006X from a WD + MS system with optically thick wind?}

 \author[label1]{X. Meng}
 \author[label1]{M. Yang}
 \address[label1]{Department of Physics and Chemistry, Henan Polytechnic
University, Jiaozuo, 454000, China, conson859@msn.com}
\author[label2]{X. Geng}
 \address[label2]{Academic publishing Center, Henan Polytechnic University,
Jiaozuo, 454000, China}


\address{}

\begin{abstract}
The single-degenerate channel is widely accepted as the
progenitors of type Ia supernovae (SNe Ia). Following the work of
Meng, Chen \& Han (2009), we reproduced the birth rate and age of
supernovae like SN 2006X by the single-degenerate model (WD + MS)
with an optically thick wind, which may imply that the progenitor
of SN 2006X is a WD + MS system.
\end{abstract}


\end{frontmatter}


\section{}
\label{sect:1} Although type Ia supernovae (SNe Ia) showed their
importance in determining cosmological parameters, e.g.
$\Omega_{\rm M}$ and $\Omega_{\Lambda}$ ((Riess et al. 1998;
Perlmutter et al. 1999), the progenitor systems of SNe Ia have not
been confidently identified yet (Hillebrandt \& Niemeyer 2000;
Leibundgut 2000). Two basic scenarios for the progenitor of SN Ia
have been discussed over the last three decades. One is a single
degenerate (SD) model (Whelan \& Iben 1973; Nomoto, Thielemann \&
Yokoi 1984), in which a CO WD increases its mass by accreting
hydrogen- or helium-rich matter from its companion, and explodes
when its mass approaches the Chandrasekhar mass limit. The
companion may be a main-sequence star (WD + MS) or a red-giant
star (WD + RG) (Yungelson et al. 1995; Li \& van den Heuvel 1997;
Hachisu et al. 1999a,b; Nomoto et al. 1999, 2003; Langer et al.
2000; Han \& Podsiadlowski 2004; Chen \& Li 2007, 2009; Han 2008;
Meng, Chen \& Han 2009; L\"{u} et al. 2009, Wang et al. 2009a, b).
An alternative is the double degenerate (DD) model (Iben \&
Tutukov 1984; Webbink 1984), in which a system consisting of two
CO WDs loses orbital angular momentum by gravitational wave
radiation and merges finally. The merger may explode if the total
mass of the system exceeds the Chandrasekhar mass limit (see the
reviews by Hillebrandt \& Niemeyer 2000 and Leibundgut 2000). In
theory, a large amount of circumstellar materials (CSM) may form
around SNe Ia via an optically thick wind for the SD model
(Hachisu et al. 1996), while there is no CSM around DD systems.
Then, a basic method to distinguish the two progenitor models is
to find the CSM around progenitor systems. The CSM may play a key
role to solve the problem of the low value of reddening ratio
(Wang 2005), which is very important for precision cosmology (Wang
et al. 2008). In addition, it is possible for the CSM to be the
origin of color excess of SNe Ia (Meng et al. 2009).

Evidence for CSM was first found in SN2002ic (Hamuy et al. 2003),
which has shown extremely pronounced hydrogen emission lines which
have been interpreted as a sign of strong interaction between
supernova ejecta and CSM. The discovery of SN2002ic may uphold the
SD model (Han \& Podsiadlowski 2006). The evidence for CSM was
also found in a normal SN Ia (SN 2006X) defined by Branch, Fisher
\& Nugent (1993) and the CSM is proposed to be from a wind from a
red-giant companion (Patat et al. 2007), while Hachisu et al.
(2008) argued a WD + MS nature for this SN Ia. Blondin et al.
(2009) found a similar signal to that of SN 2006X in another SNe
Ia (SN 1999cl) and their results indicated that the supernovae
like SN 2006X are rare objects ($2/31\sim6\%$). Recently, the
third example like SN 2006X (SN 2007le) was reported (Simon et al.
2009), and then the ratio of 2006X-like supernova is increased to
$3/32\sim9\%$.

In theory, SNe Ia can explode during the optically thick wind
phase or after the optically thick wind while in stable or
unstable hydrogen-burning phase (Han \& Podsiadlowski 2004). The
materials lost as the optically thick wind form CSM (Hachisu et
al. 2008; Meng et al. 2009). If a SN Ia explodes after the
optically thick wind, the CSM have been dispersed too thin to be
detected immediately after the SN Ia explosion (Hachisu et al.
2008; Meng, Yang \& Geng 2009). If a SNe Ia explodes during the
optically thick wind phase, the materials lost from system form
CSM very near the SN Ia, which may show the signal similar to SN
2006X (Hachisu et al. 2008; Meng, Yang \& Geng 2009). Recently,
Meng, Chen \& Han (2009) performed binary stellar evolution
calculations for more than 25,000 close WD + MS binary systems
with various metallicities. In their works, the prescription of
Hachisu et al. (1999a) for the accretion efficiency of
hydrogen-rich material was adopted by assuming an optically thick
wind (Hachisu et al. 1996), and then their works provide a
possibility to check whether the SD model with optically thick
wind can explain the birth rate of supernovae like SN 2006X. The
purpose of this paper is to check the possibility by a binary
population synthesis (BPS) approach.

In section \ref{sect:2}, we simply describe our method. We show
the results in section \ref{sect:3} and give discussions and
conclusions in section \ref{sect:4}.
\section{Method}
\label{sect:2}
\subsection{optically thick wind}
\label{sect:2.1} Meng, Chen \& Han (2009) studied the WD+MS system
with optically thick wind. In their studies, they adopted the
prescription of Hachisu et al. (1999a) on WDs accreting
hydrogen-rich material from their companions instead of solving
stellar structure equations of a WD. In a WD + MS channel, the
companion fills its Roche lobe at MS or during HG, and transfers
material onto the WD. If the mass-transfer rate, $|\dot{M}_{\rm
2}|$, exceeds a critical value, $\dot{M}_{\rm cr}$, they assumed
that the accreted hydrogen steadily burns on the surface of WD and
that the hydrogen-rich material is converted into helium at the
rate of $\dot{M}_{\rm cr}$. The unprocessed matter is assumed to
be lost from the system as an optically thick wind at a rate of
$\dot{M}_{\rm wind}=|\dot{M}_{\rm 2}|-\dot{M}_{\rm cr}$ (Hachisu
et al. 1996). The hydrogen-rich material lost as the optically
thick wind may exist as CSM. If the WD explodes at the wind phase,
the CSM locates near the SN Ia, which may show the signal similar
to SN 2006X (Hachisu et al. 2008; Meng, Yang \& Geng 2009). In
this paper, we assume that if SN Ia explodes during the optically
thick wind phase, it is a supernova like SN 2006X.

\subsection{method}
\label{sect:2.2}

We use the rapid binary evolution code developed by Hurley et al.
(2000, 2002) to study the birth rate of supernovae like SN 2006X.
To investigate the birth rate, we followed the evolution of
$10^{\rm 7}$ binaries. The primordial binary samples are generated
in a Monte Carlo way and a circular orbit is assumed for all
binaries. The basic parameters for the simulations are the same as
that in Meng, Chen \& Han (2009). In theory, when the initial WD
mass in a binary system is smaller than $0.9M_{\odot}$, the WD
never explodes at the optically thick wind phase (Hachisu et al.
2008; Meng, Yang \& Geng 2009; Meng, Chen \& Han 2009). Meng, Chen
\& Han (2009) have shown the parameter spaces for the supernovae
exploding at the wind phase (see Fig. 2 in Meng, Chen \& Han
2009), and we assume that these supernovae are those like SN
2006X. For our binary population synthesis study, we assume that
if the initial orbital-period, $P_{\rm orb}^{\rm i}$, and the
initial secondary mass, $M_{\rm 2}^{\rm i}$, of a WD + MS system
is located in the appropriate regions in the ($\log P^{\rm i},
M_{\rm 2}^{\rm i}$) plane for supernovae like SN 2006X at the
onset of RLOF, a 2006X-like supernova is produced.

For our BPS study, common-envelope (CE) ejection efficiency
($\alpha_{\rm CE}$, the fraction of the released orbital energy
used to eject the CE) is a very important parameter, and in this
paper, we set $\alpha_{\rm CE}$ to be 1.0 or 3.0. Two cases for
star formation history are checked, i.e. a single starburst and a
constant star formation rate ($SFR$) over the last 15 Gyr. For the
constant star formation rate, we assume that one binary with
primary larger than $0.8 M_{\odot}$ is born in the Galaxy each
year (see Iben \& Tutukov 1984; Han, Podsiadlowski \& Eggleton
1995; Hurley et al. 2002). From this calibration, we can get
$SFR=5 M_{\odot}$ ${\rm yr^{-1}}$ (see also Willems \& Kolb 2004).

\section{The Results of Binary Population Synthesis}\label{sect:3}

\begin{figure}
\includegraphics[angle=270,scale=.50]{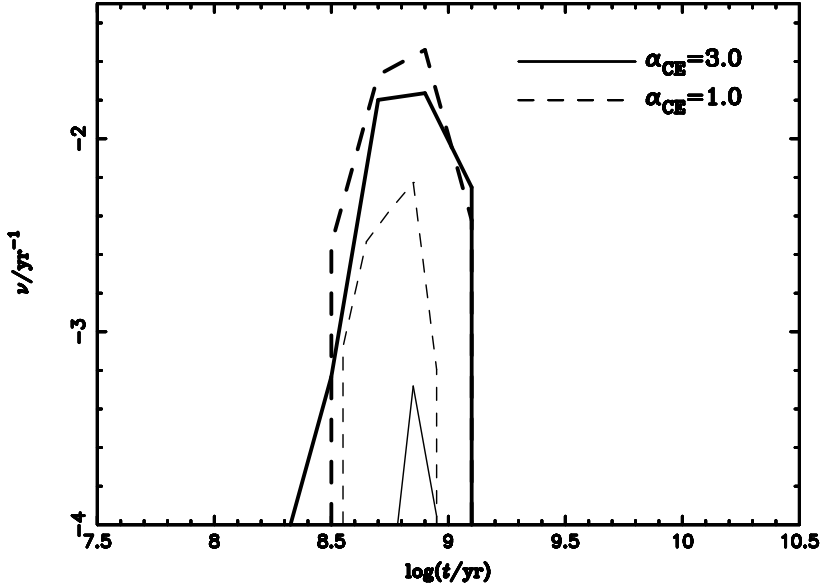}
\caption{The evolution of the birth rates of SNe Ia for a single
starburst of $10^{\rm 11}M_{\odot}$ with different $\alpha_{\rm
CE}$. The thick lines are for all the SNe Ia, while the thin lines
are for those like SN 2006X.}\label{single}
\end{figure}

\begin{figure}
\includegraphics[angle=270,scale=.50]{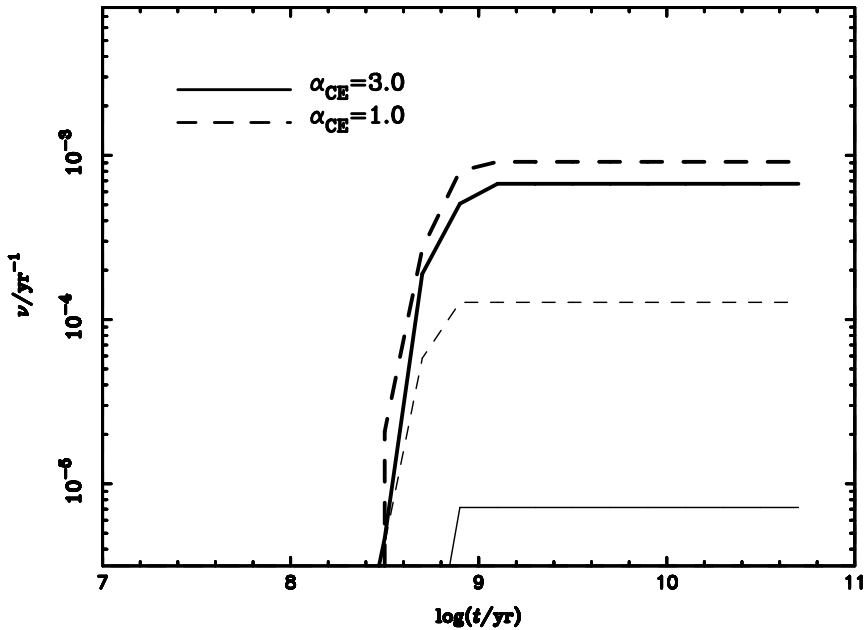}
\caption{Similar to Fig. \ref{single} but for a constant star
formation rate ($5M_{\odot}{\rm yr}^{\rm -1}$).}\label{const}
\end{figure}

In Figs. \ref{single} and \ref{const}, we show the evolution of
the birth rates for all SNe Ia and supernovae like SN 2006X with a
single starburst and a constant star formation rate, respectively.
From the two figures, we can see that SNe similar to SN 2006X are
a relatively rare subclass of SNe Ia: 1 - 14 in 100 SNe Ia can be
of this type, which depends on the $\alpha_{\rm CE}$. The
observational results in Blondin et al. (2009) indicated that out
of 100 SNe Ia, about 6 supernovae should belong to the subclass of
2006X-like objects. The probability is located in the range given
in this paper, even that SN 2007le is included.

In the figures, we can see that $\alpha_{\rm CE}$ affects the
birth rate significantly. High $\alpha_{\rm CE}$ results in a
lower birth rate. This result results from the influence of
$\alpha_{\rm CE}$ on the channels forming WD + MS systems. There
are three channels to form WD + MS systems: He star channel, EAGB
channel and TPAGB channel. The classification is based on the
evolutionary phase of the primordial primary at the onset of the
first Roche Lobe overflow (RLOF), which means that the primordial
primary is in HG or on RGB (i.e. case B evolution defined by
Kippenhahn \& Weigert (1967), in early asymptotic giant branch
stage (EAGB) (i.e. helium is exhausted in the core, while thermal
pulses have not yet started) or at the thermal pulsing AGB (TPAGB)
stage for He star, EAGB and TPAGB channels, respectively (see
Meng, Chen \& Han 2009 for the channels in details). Because of
the low binding energy of the common envelope and a long
primordial orbital period, $\alpha_{\rm CE}$ has a remarkable
influence on CO + WD systems from the TPAGB channel. Generally, if
a CE can be ejected, a low $\alpha_{\rm CE}$ produces a shorter
orbital-period WD + MS system, which is more likely to fulfill the
conditions for SNe Ia. Therefore, we see obvious contributions
from the TPAGB channel when $\alpha_{\rm CE}=1.0$, but no
contribution from this channel when $\alpha_{\rm CE}=3.0$. This is
reason why a high $\alpha_{\rm CE}$ leads to low birth rate.

From Fig. \ref{single}, we can see that if SN 2006X is originated
from a WD + MS system as suggested in this paper, the delayed time
of 2006X-like object is constrained in a very narrow range, i.e.
$0.3\sim1{\rm Gyr}$. This provides a rigorous constraint on the
age of the object like SN 2006X. We will discuss it in the next
section.

\section{Discussion and Conclusions}\label{sect:4}
\subsection{the age of SN 2006X}
\label{sect:4.1}

SN 2006X is the first case that show a variable Na I D line in its
spectral, which clearly shows a signal of CSM (Patat et al. 2007).
Patat et al. (2007) also noticed a relatively low expansion
velocity of the CSM ($\sim50$ ${\rm km/s}$). The low expansion
velocity seems to imply that the progenitor of SN 2006X belongs to
a WD + RG systems. However, Hachisu et al. (2008) showed that
under the assumption of the optically thick wind, the WD + MS
system also may produce the low-velocity CSM. In this paper, we
show that if SN 2006X is from a WD + MS system with the optically
thick wind, the age of its progenitor is smaller than 1 Gyr, which
implies that there is star formation during the recent 1 Gyr in
its host galaxy. The host galaxy of SN 2006X, NGC 4321 (M100), is
a well-studied spiral galaxy (e.g. Ho et al. 1997) and SN 2006X
locates near one of its arms (Wang et al. 2008). As one of the
largest spiral galaxies in the Virgo cluster, it produced SNe
1901B, 1914A, 1959E, 1979C, and 2006X in roughly the last century
and shows significant signal of star formation at present (Kanpen
et al. 1993, 1996). Recently, Blondin et al. (2009) found a
similar signal to that of SN 2006X in another SNe Ia (SN 1999cl)
in archives. Its host galaxy, NGC4501 (M88), is also a spiral
galaxy and SN 1999cl locates at one of its arms (Krisciunas et al.
2000). During the last 1 Gyr, the star formation is significant in
the galaxy (Wong \& Blitz 2002). The host galaxy of SN 2007le (NGC
7721) is a Sc galaxy, and a recent star formation is also expected
(Iglesias-P\'{a}ramo et al. 2006; Simon et al. 2009). So, all the
supernovae fulfill the age constraint from WD + MS system.
\subsection{Progenitor system}
\label{sect:4.2}

Following the study of Meng, Chen \& Han (2009), in this paper, we
show the evolution of birth rate of supernovae like SN 2006X by
assuming that if a SN Ia explodes at the optically thick wind
phase, it is a 2006X-like supernova. Based on the assumption, we
can reproduce the birth rate of 2006X-like objects. Please keep in
mind that we only consider the case of WD + MS channel in this
paper. However, the progenitor of SN 2006X is still an open
question. Patat et al. (2007) suggested that the progenitor of SN
2006X belongs to WD + RG system such a RS Oph based on the
velocity of absorptions line of Na I D. L\"{u} et al. (2009)
designed a WD + RG model with an aspherical stellar wind and
equatorial disk, and then they may explain some properties of SN
2006X. However, the birth rate of 2006X-like objects obtained in
L\"{u} et al. (2009) is too low (less than 1\%) to compare with
that observed. Based on the assumption of the optically thick wind
and a mass-stripping effect, Hachisu et al. (2008) argued that the
progenitor of SN 2006X should be a WD + MS system and they also
can well explain the properties of SN 2006X. Although the
treatment of binary evolution in Meng, Chen \& Han (2009) is
different from that in Hachisu et al. (2008), Meng, Chen \& Han
(2009) obtained similar results for the SNe Ia exploding at the
optically thick wind phase, e.g. there is no supernova explosion
at the optically thick wind phase when the initial mass of CO WD
is smaller than 0.9 $M_{\odot}$. So, considering the results in
this paper, all the properties of SN 2006X can be explained by the
WD + MS channel with the optically thick wind, including its birth
rate and its age. Then, we suggest that the progenitor of
supernovae like SN 2006X is WD + MS system, not WD + RG system.



\end{document}